\documentclass[twocolumn,showpacs,preprintnumbers,amsmath,amssymb,prl,aps,10pt,superscriptaddress,nofootinbib,reprint]{revtex4-1}

\usepackage{graphicx}
\usepackage{dcolumn}
\usepackage{bm}
\usepackage{dcolumn}
\usepackage{amsmath}
\usepackage{amssymb}
\usepackage{longtable}
\usepackage{array}
\usepackage[version=3]{mhchem}
\usepackage{epic}
\usepackage{epstopdf}
\usepackage{verbatim} 

\usepackage[usenames,dvipsnames,svgnames,table]{xcolor}

\newcolumntype{C}[1]{>{\centering}m{#1}}
\newcommand{\etal}{\textit{et al.}}
\newcommand{\bv}[1]{{\boldsymbol #1}}

\newcommand{\muB}{~\mu _\text{B}}
\newcommand{\rA}{$\Delta \rho$}

\begin{document}


\title{In-plane anisotropic magnetoresistance in antiferromagnetic \\ \boldmath Ba(Fe$_{1-x}$Co$_x$)$_2$As$_2$, (Ba$_{1-x}$K$_x$)Fe$_2$As$_2$ and Ba(Fe$_{1-x}$Ru$_x$)$_2$As$_2$ \unboldmath}

\author{Gerald Derondeau}    \email{gerald.derondeau@cup.uni-muenchen.de}
\affiliation{%
  Department  Chemie,  Physikalische  Chemie,  Universit\"at  M\"unchen,
  Butenandtstr.  5-13, 81377 M\"unchen, Germany\\}

\author{J\'an Min\'ar}
\affiliation{%
  Department  Chemie,  Physikalische  Chemie,  Universit\"at  M\"unchen,
  Butenandtstr. 5-13, 81377 M\"unchen, Germany\\}
\affiliation{%
  NewTechnologies-Research Center, University of West Bohemia, Pilsen, Czech Republic\\}

\author{Sebastian Wimmer}
\affiliation{%
  Department  Chemie,  Physikalische  Chemie,  Universit\"at  M\"unchen,
  Butenandtstr. 5-13, 81377 M\"unchen, Germany\\}

\author{Hubert Ebert}
\affiliation{%
  Department  Chemie,  Physikalische  Chemie,  Universit\"at  M\"unchen,
  Butenandtstr. 5-13, 81377 M\"unchen, Germany\\}

\date{\today}

\begin{abstract}
Using the Kubo-Greenwood formalism the resistivity anisotropy for electron doped Ba(Fe$_{1-x}$Co$_x$)$_2$As$_2$, hole doped (Ba$_{1-x}$K$_x$)Fe$_2$As$_2$ and isovalently doped Ba(Fe$_{1-x}$Ru$_x$)$_2$As$_2$ in their antiferromagnetic state has been calculated in order to clarify the origin of this important phenomenon. The results show good agreement with experiment for all cases without considering impurity states extending over several unit cells or temperature induced spin fluctuations. From this it is concluded that the resistivity anisotropy at low temperatures is primarily caused by an in-plane anisotropic magnetoresistance. 
Accounting for the band dispersion with respect to $k_z$ is however mandatory to explain the results, showing the importance of the three-dimensional character of the electronic structure for the iron pnictides. Furthermore, it is shown that the counterintuitive sign of the resistivity anisotropy is no fundamental property but just a peculiarity of the anisotropic band structure.
\end{abstract}
\maketitle


The iron pnictide superconductors are one of most important examples for unconventioanl superconductivity, with the physical interest of the last few years mainly focused on their strong in-plane anisotropy which emerges for various physical properties \cite{CAG+10,TBK+10,INL+13,YLC+11,ZAY+09,NIT+12}. This anisotropy is prominent in electrical transport \cite{CAG+10,TBK+10,YWW+11,CKAF12,INL+13,INL+13a,BTF+13,KF14} but also shows up in angle-resolved photoemission spectroscopy (ARPES) \cite{YLC+11,KOK+11}, neutron diffraction \cite{ZAY+09} or optical spectroscopy \cite{NIT+12,DLP+11}. 
Of special interest is the in-plane resistivity anisotropy (\rA~$= \rho_a - \rho_b$) for the \ce{BaFe2As2} family as it seems to involve several physical phenomena. Notably, the higher resistivity was observed along the shorter $b$ axis ($\rho_b$) which corresponds to ferromagnetically coupled chains, while the lower resistivity was found along the $a$ axis ($\rho_a$) along which the spins couple antiferromagnetically \cite{TBK+10,CAG+10}. A quite different behavior is reported for different kinds of substitution in \ce{BaFe2As2} \cite{KCR+11,INL+13a,BTF+13,LMI+15}.  Compared to electron doped \ce{Ba(Fe_{1-$x$}Co_{$x$})2As2} \cite{CAG+10,INL+13} a smaller resistivity and anisotropy is seen in isovalent \ce{Ba(Fe_{1-$x$}Ru_{$x$})2As2} \cite{LMI+15} and an almost negligible resistivity and possibly inverted anisotropy is reported for hole doped \ce{(Ba_{1-$x$}K_{$x$})Fe2As2} \cite{YWW+11,BTF+13}. 

Obviously, it turned out to be non-trivial to find a theoretical explanation which can account for all these findings. There is recent theoretical work by several groups based on model Hamiltonian approaches \cite{FAS+11,GPW+14,GHA14,WGA+15,SPKT14,BBT14,BBT14a} and little numerical work based on Monte Carlo simulations \cite{LAS+12}.
However, the origin of \rA~is still under discussion. For example Kuo and Fisher \cite{KF14} ascribe \rA~exclusively to contributions of the intrinsic anisotropic band structure. Another approach is to assume extrinsic defect states which extend over several unit cells \cite{INL+13,LMI+15}. These can be explained theoretically \cite{GHA14,GPW+14,WGA+15} and monitored by scanning tunneling microscope (STM) experiments which see local anisotropic defect states with a size of approximately 22~\AA~ \cite{CAL+10,ACM+13,RAA+14}.
What is missing so far is theoretical work that is not based on a model Hamiltonian but on a parameter-free application of density functional theory (DFT) or a comparable scheme, showing whether and how the intrinsic band structure can explain the observed resistivity anisotropy \rA.  

With this work we present a first principles DFT-based study on \rA~in three exemplary iron pnictides: electron doped \ce{Ba(Fe_{1-$x$}Co_{$x$})2As2} (Co-122), hole doped \ce{(Ba_{1-$x$}K_{$x$})Fe2As2} (K-122) and isovalent doped \ce{Ba(Fe_{1-$x$}Ru_{$x$})2As2} (Ru-122). In order to reliably account for the disorder induced by substitution we use the Korringa-Kohn-Rostoker-Green function (KKR-GF) approach together with the coherent potential approximation (CPA) alloy theory, which proved already to be a very powerful and accurate tool to study the electronic structure of the substitutional iron pnictides \cite{DPM+14,KJ14,DBEM16}. Access to the longitudinal resistivity is given by the Kubo-Greenwood equation \cite{KLSE11,KCE15}, which allows for a direct comparison with experiment. 
All calculations have been performed self-consistently and fully relativistically within the four component Dirac formalism based on the local density approximation (LDA) \cite{EKM11,SPR-KKR6.3_2}. 
The applied theory thus accounts for all effects of the intrinsic band structure and of disorder induced impurity scattering in terms of the CPA. On the other hand, it does not account for anisotropic impurity states extending over several unit cells or for temperature induced spin fluctuations. More details are found in the Supplemental Material.\cite{Supp}

%

\begin{figure}[tb]
{\includegraphics[clip]{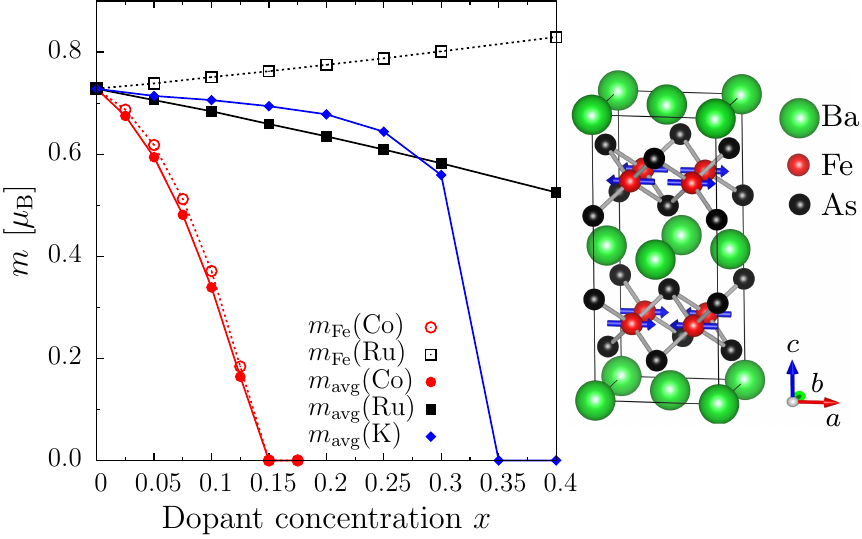}} 
\caption{(Left) Magnetic moments for Co-122 (red), K-122 (blue) and Ru-122 (black) depending on the dopant concentration $x$. The solid lines correspond to $m_\text{avg}$, meaning the substitutional dependent average, while the dashed lines correspond to the pure Fe moment $m_\text{Fe}$. (Right) Orthorhombic unit cell with the experimental magnetic configuration.\vspace*{-0.2cm}}\label{Fig_MagnMomComp}
\end{figure}

The magnetic properties of the investigated compounds are summarized in Fig.~\ref{Fig_MagnMomComp} together with the  orthorhombic unit cell in its experimental magnetic configuration. The obtained magnetic moment of undoped \ce{BaFe2As2} is $0.73\muB$ and thus in good agreement with experimental data ($0.87\muB$ \cite{HQB+08}). Using the CPA allows for a site resolved investigation of the magnetic moments, thus, the solid lines correspond to the average magnetic moment while the dashed lines give the Fe contribution only. The collapse of long range antiferromagnetic order at the critical concentration $x_\text{crit}$ for Co-122 and K-122 is in reasonable agreement with experiment, although it is slightly shifted to higher $x$ values when compared with experiment (Co-122: $x_\text{crit}^\text{exp} \approx 0.07$ \cite{LCA+09}; K-122: $x_\text{crit}^\text{exp} \approx 0.25$ \cite{RPTJ08}). Yet, for isovalent doped Ru-122 the average magnetic moments decrease due to the increasing Ru concentration which has a low magnetic moment in the order of $0.07\muB$ (not shown in Fig.~\ref{Fig_MagnMomComp}), however, the individual Fe magnetic moments surprisingly increase. In the literature magnetic dilution was already discussed as the main driving force for the magnetic breakdown in Ru-122 \cite{DLF+11,DHR+13}, yet in the investigated regime of $x$ the magnetic order did not disappear, i.e. $x_\text{crit}$ should be higher than 0.4 (see Fig.~\ref{Fig_MagnMomComp}).

\begin{figure*}[tbh]
{\includegraphics[clip]{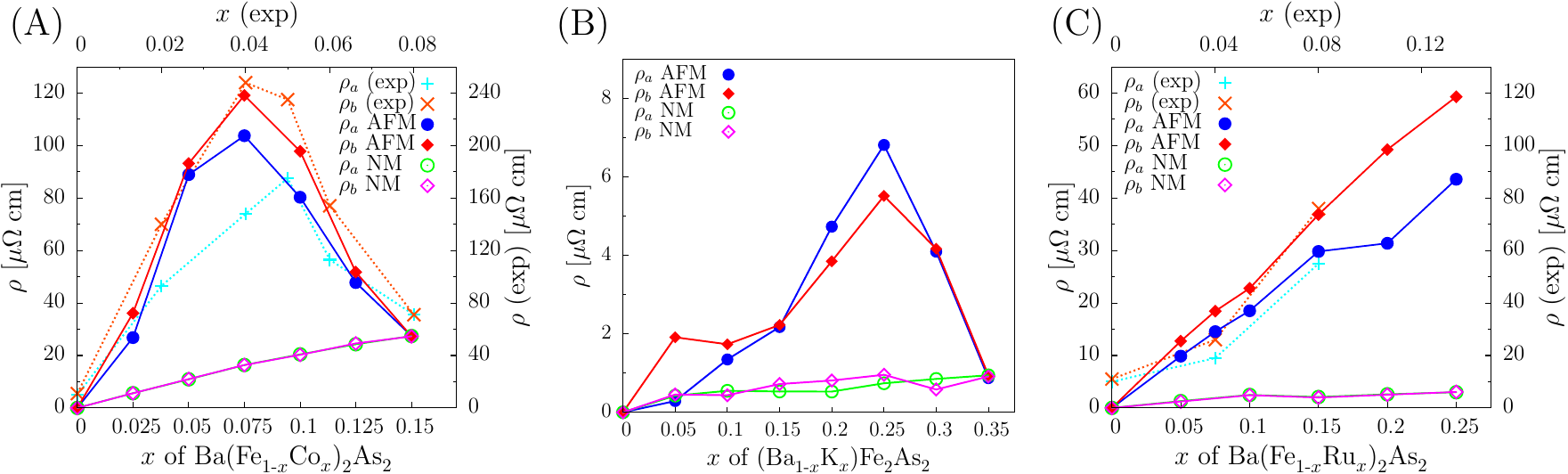}} 
\caption{In-plane resistivity calculated for (A) Co-122, for (B) K-122 and for (C) Ru-122 up to $x_\text{crit}$ as a function of the concentration $x$. The blue and red curves correspond to the antiferromagnetic (AFM) state, while the green and purple lines correspond to the same orthorhombic lattice but for the nonmagnetic (NM) case. The pluses and crosses are experimental data \cite{INL+13,LMI+15} with the corresponding concentration $x$ (exp) and resistivity $\rho$ (exp) axes given at the top and right of the figure.\vspace*{-0.2cm}}\label{Fig_ResComp}
\end{figure*}

Accounting for the magnetic configuration the in-plane longitudinal resistivity was calculated, with the main results shown in Fig.~\ref{Fig_ResComp}. First consider Co-122 in (A), where the resistivity of the antiferromagnetic (AFM) configuration has a dome-like variation with increasing dopant concentration $x$ until the collapse of the AFM order takes place at $x_\text{crit}$. Note that the resistivity $\rho_b$ AFM along the ferromagnetic $b$ axis (red) is always larger than the resistivity $\rho_a$ AFM along the antiferromagnetic $a$ axis  (blue) with a maximum of the anisotropy roughly at $x_\text{crit}$/2.
This behavior is in full agreement with experiment \cite{CAG+10,INL+13}. 
In Fig.~\ref{Fig_ResComp} (A) we also give the experimental data of Ishida \etal \cite{INL+13} for $\rho_a$ and $\rho_b$ using specifically adapted scales to the right and top of the figure. A direct comparison of the theoretical and experimental data seems nevertheless justified for two reasons.
First of all, the self-consistently calculated collapse of the AFM order takes place at a higher $x_\text{crit}$ compared to experiment, thus one should account for that by adjusting the experimental and theoretical $x$ axis in a way that the breakdown of magnetic order coincides. Secondly, it is known that the resistivity decreases significantly for annealed crystals, indicating a strong contribution of crystal defects \cite{INL+13,INL+13a}. Thus it is expected to find the calculated resistivity to be lower than the experimental one, consequently the axis to the right for the experimental resistivity is adjusted by a factor of 2. Most important are the order of magnitude and the dependency of \rA~on the doping, which are rather well reproduced by the presented calculations. For comparison we show also the calculated resistivity of the same orthorhombic crystal but for the nonmagnetic (NM) state, using green ($a$ axis) and purple ($b$ axis) lines in Fig.~\ref{Fig_ResComp}. Notably, the resistivity of the orthorhombic NM Co-122 is almost by a factor of 10 reduced compared to the AFM case and it shows neither a dome-like behavior nor a significant anisotropy due to the lattice distortion. This unambiguously shows that the lattice has a negligible contribution to the anisotropic behavior compared to the magnetic structure, as was stressed before in Refs \cite{DPM+14,DBEM16}.

Next, consider K-122 in Fig.~\ref{Fig_ResComp} (B), where we show no experimental data because \rA~is reported to be almost immeasurably small with even possible sign changes \cite{YWW+11,BTF+13}. 
Indeed, the calculations find an in-plane resistivity always below 8\;$\mu \Omega$\;cm which is reduced by a factor of 15 compared to Co-122. Replacing Fe with Co or Ru affects the $d$-electronic states which are dominating at the Fermi level ($E_F$) for the iron pnictides. Thus, disorder introduced for $sp$-elements like Ba or K hardly affects the resistivity of the compound. The calculations further support the possibility of sign changes in \rA~with $x$.

Finally, consider Ru-122 in Fig.~\ref{Fig_ResComp} (C), with recent experimental data taken from Liu \etal\;\cite{LMI+15} shown as for the Co-122 case with axes at the top and on the right adjusted the same way. The sign of \rA~is the same as for Co-122, however, the anisotropy does not disappear as the magnetic order in Ru-122 (see Fig.~\ref{Fig_MagnMomComp}) also does not disappear for the investigated regime of substitution. In addition, one has to note that for comparable dopant concentrations the resistivity in Ru-122 is reduced compared to Co-122, as it was also observed in experiment \cite{LMI+15}. Thus, again the trends and the order of magnitude for the resistivity are in reasonable agreement with recent experimental data. 

In summary, in all three representative iron pnictides it was possible to reproduce the qualitative behavior of \rA~in the antiferromagnetic phase based on a LDA approach without considering spatially extended impurity states or spin fluctuations. Thus, the resistivity anisotropy of the iron pnictides at low temperatures can be well understood in terms of an in-plane anisotropic magnetoresistance (AMR) \cite{MFF+14}. This leads to the question how the anisotropic band structure can influence this AMR. The idea of linking \rA~to the band structure as seen e.g. by ARPES was already proposed by Yi \etal\;\cite{YLC+11}, however, corresponding explanations were hardly successful so far \cite{LMI+15}. The main reason for this apparent incompatibility is due to the fact that ARPES is a surface sensitive method. For the iron pnictides, however, it is absolutely mandatory to account for the dispersion of bands with $k_z$ as was also recently stressed in the context of the effective mass enhancement seen in ARPES \cite{DBB+16}. Accounting for the $k_z$ dispersion of the anisotropic bands will thus allow for more meaningful results and an at least qualitative understanding of the observed trends.

\begin{figure*}[tb]
{\includegraphics[clip]{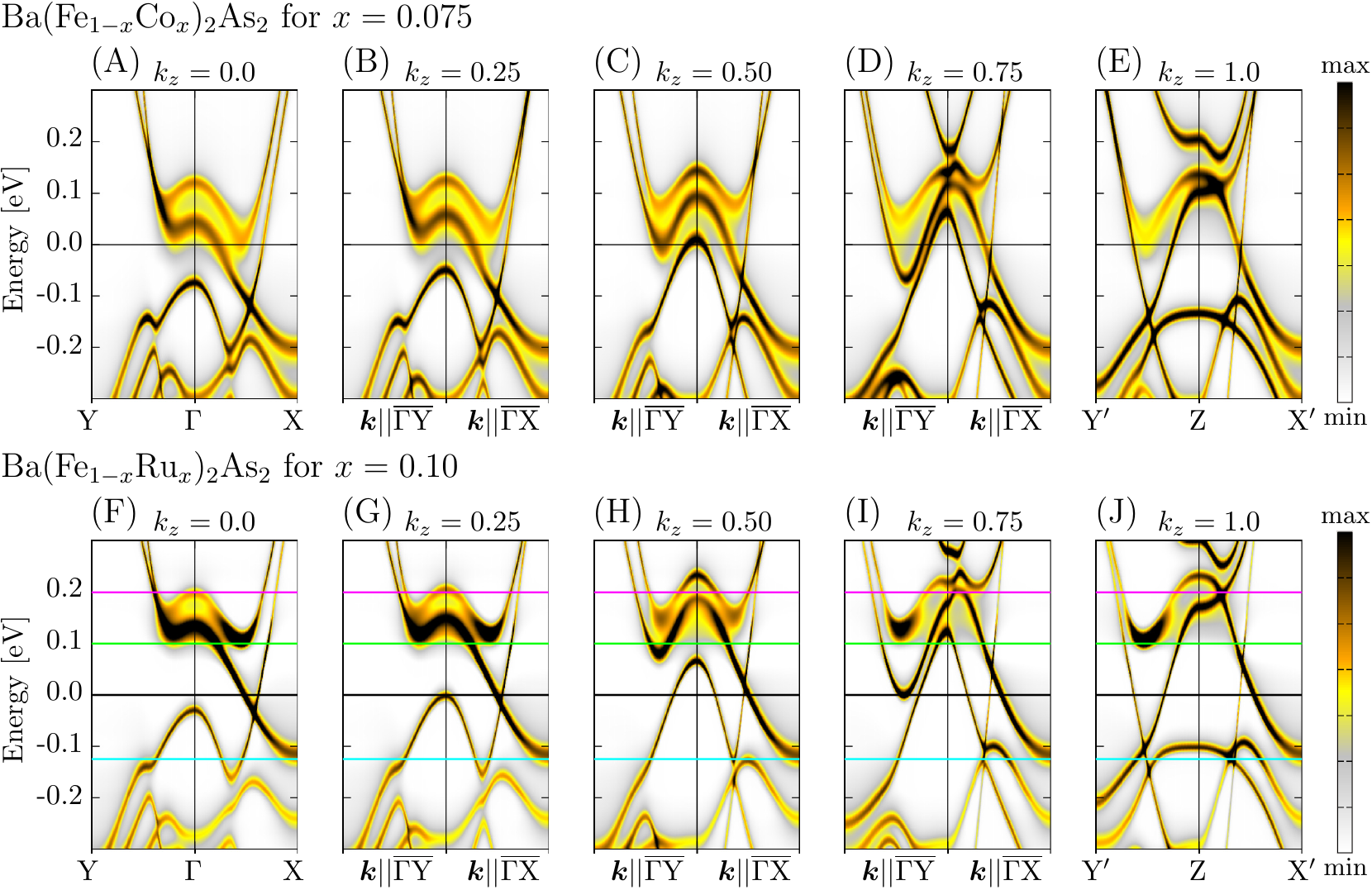}} 
\caption{Bloch spectral functions for different values of $k_z$ in units of $\frac{2\pi}{c}$, with $\overline{\Gamma\mathrm{Y}}$ and $\overline{\Gamma\mathrm{X}}$ corresponding to paths parallel to the $b$ and $a$ axes, respectively. The panels (A - E) and (F - J) show results for Co-122 with $x=0.075$ and Ru-122 with $x=0.10$, respectively, with the colored cuts corresponding to the cuts in Fig.~\ref{Fig_EF}.\vspace*{-0.2cm}}\label{Fig_Band}
\end{figure*}

\begin{figure}[b]
{\includegraphics[clip]{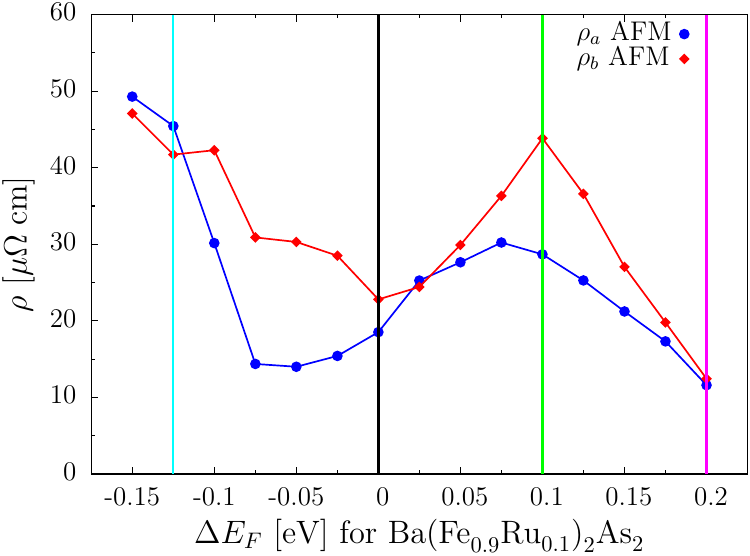}} 
\caption{Change in $\rho_{a(b)}$ for Ru-122 with $x=0.1$ depending on an artificial shift $\Delta E_F$ of the Fermi energy. The cyan, black, green and purple cuts correspond to the cuts in the band structures in Fig.~\ref{Fig_Band}.\vspace*{-0.2cm}}\label{Fig_EF}
\end{figure}

Discussing the presented resistivity data, one should first note that the resistivity anisotropy is also observed for the parent compound \ce{BaFe2As2}, although it is reported to be small after annealing \cite{TBK+10,INL+13a}. Still, the Lifshitz transition \cite{LKF+10}, i.e. the topological change of the Fermi surface when going from the nonmagnetic to the antiferromagnetic state, should influence the resistivity behavior. Indeed, the formerly almost isotropic bands in the NM state undergo a significant band splitting resulting in a considerable anisotropy after the Lifshitz transition to the AFM state \cite{LKF+10,DPM+14,DBEM16}. When accounting for the $k_z$ dispersion, one can qualitatively understand the higher resistivity along $b$ in terms of a hybridization of hole and electron pockets seen also in ARPES \cite{LKF+10}. This leads to the formation of a band gap and of minima from electron pockets at the Fermi level only along the $b$ axis. This Lifshitz transition of \ce{BaFe2As2} and the implications for \rA~are discussed in more detail in the Supplemental Material.\cite{Supp} 

A qualitative discussion of \rA~for the investigated compounds will be exemplarily done for Co-122 with $x=0.075$, having the highest value of \rA~for Co substitution and for Ru-122 with $x=0.10$ as an example of isovalent doping. The band structures are shown for different $k_z$ values in Fig.~\ref{Fig_Band} with the paths $\overline{\Gamma\mathrm{Y}}$ and $\overline{\Gamma\mathrm{X}}$ corresponding to directions in $\bv k$-space parallel to the $b$ and $a$ axes, respectively. First consider the Co-122 case in panels (A - E), where a strong dependence of the bands on $k_z$ is obvious. Of most interest are the W- or S-shaped bands (W-band: two minima along both directions $\overline{\Gamma\mathrm{Y}}$ and $\overline{\Gamma\mathrm{X}}$; S-band: only one minimum along $\overline{\Gamma\mathrm{Y}}$) which are a result of the previously discussed hybridization with significant band broadening due to the disorder. The highest \rA~is observed for $x =0.075$ in Co-122 because for exactly this concentration most of these W- and S-bands have their minimum precisely at $E_F$. Approaching these extrema, the slope of the bands is decreasing, leading to a decrease of the quasiparticle velocities and thus to an increase of the resistivity. Additionally, the apparently strong disorder-induced broadening of the W- and S-bands leads to a decreased quasiparticle lifetime which again increases the resistivity. It would be difficult to explain \rA~based on their contribution at $E_F$ in Co-122 for some values of $k_z$, e.g. for Fig.~\ref{Fig_Band} (A) $k_z = 0.0$, (B) 0.25 and (D) 0.75. However, for (C) $k_z = 0.5$ and (E) 1.0 the contribution at $E_F$ is clearly anisotropic with a higher impact along the $\overline{\Gamma\mathrm{Y}}$ path, leading to a higher resistivity $\rho_b$ compared to $\rho_a$. 

Analogously, \rA~can be explained for the Ru-122 compound with $x=0.1$, but now the strong anisotropic contribution can be seen for Fig.~\ref{Fig_Band} (F) $k_z = 0.0$ and (I) 0.75, where one has either a band gap along $\overline{\Gamma\mathrm{Y}}$, or the minimum of a S-shaped band at $E_F$.  
The Ru-122 system is an interesting prototype system for isovalent doping because the Ru substitution induces disorder but $E_F$ is only marginally changed. The crucial impact of the anisotropic band structure can be easily shown by artificially changing the Fermi energy by $\Delta E_F$ and recalculating the resistivity $\rho_{a (b)}$ for Ba(Fe$_{0.9}$Ru$_{0.1}$)$_2$As$_2$. The results are presented in Fig.~\ref{Fig_EF} where some exemplary values are highlighted by the cyan, black, green and purple lines. The corresponding energies are indicated in Fig.~\ref{Fig_Band} (F - J) for comparison. 
It becomes immediately obvious, that the magnitude and the sign of $\Delta \rho = \rho_a -\rho_b$ depends strongly on the chosen energy. For $\Delta E_F = -0.125$\;eV (cyan line) one arrives at the minimum of a band at X, whose impact is strong enough to change the sign of \rA, meaning a higher resistivity $\rho_a$ compared to $\rho_b$. This is comparable to K-122, where the hole doping induces this energy shift and explains the possible sign change in \rA. Increasing the energy to $\Delta E_F = 0.10$\;eV (green line) significantly increases \rA~that gets comparable to the Co-122 case for $x=0.075$ in Fig.~\ref{Fig_Band} (A - E). Further increasing the shift to $\Delta E_F = 0.20$\;eV (purple line) moves again away from this peculiar band situation, decreasing \rA~as it is the case for overdoped Co-122 with $x > 0.1$. 

Altogether, we have shown that \rA~in the low temperature phase of the iron pnictides can be explained in terms of an in-plane AMR.  Additional contributions of extended impurity states obviously cannot be ruled out. Furthermore, we do not disagree with work considering the impact of higher temperatures in the nematic phase \cite{GHA14,GPW+14,WGA+15, BBT14}, rather we provide new insights from first-principles calculations on real materials. These disproved in particular recent suggestions that the anisotropic band structure has no impact on \rA~\cite{LMI+15} and highlighted the importance of the dispersion with $k_z$.

In conclusion, this work is the first to present first principles transport calculations of the resistivity anisotropy in the low temperature antiferromagnetic phase of the iron pnictide superconductors. We show for three exemplary systems resistivity values in good agreement with experiment. It turned out that it is sufficient to discuss \rA~in the antiferromagnetic phase in terms of an anisotropic magnetoresistance. This AMR can be tuned by either disorder scattering from impurities or by an energy shift due to respective electron or hole doping.
Most important, it was mandatory to account for the $k_z$ dispersion in order to understand the results based on the anisotropic band structure, showing the crucial impact of the three-dimensional character of the electronic structure for the iron pnictides which future studies have to account for. Furthermore, resistivity calculations for varying energy reveal that the sign of \rA~is no fundamental property of the pnictides but rather a peculiarity of the anisotropic band structure.

\begin{acknowledgments}
We thank Ilja Turek for valuable discussions. We acknowledge financial support within the research group FOR 1346, within priority program SPP 1538 and from CENTEM PLUS (L01402). 
\end{acknowledgments}
%

\end{document}


\setstretch{1.4}

\title{{\huge Supplemental Material for} \\\vspace{1cm} {In-plane anisotropic magnetoresistance in antiferromagnetic \\ \boldmath Ba(Fe$_{1-x}$Co$_x$)$_2$As$_2$, (Ba$_{1-x}$K$_x$)Fe$_2$As$_2$ and Ba(Fe$_{1-x}$Ru$_x$)$_2$As$_2$ \unboldmath}}

\author{Gerald Derondeau}    \email{gerald.derondeau@cup.uni-muenchen.de}
\affiliation{%
  Department  Chemie,  Physikalische  Chemie,  Universit\"at  M\"unchen,
  Butenandtstr.  5-13, 81377 M\"unchen, Germany\\}

\author{J\'an Min\'ar}
\affiliation{%
  Department  Chemie,  Physikalische  Chemie,  Universit\"at  M\"unchen,
  Butenandtstr. 5-13, 81377 M\"unchen, Germany\\}
\affiliation{%
  NewTechnologies-Research Center, University of West Bohemia, Pilsen, Czech Republic\\}

\author{Sebastian Wimmer}
\affiliation{%
  Department  Chemie,  Physikalische  Chemie,  Universit\"at  M\"unchen,
  Butenandtstr. 5-13, 81377 M\"unchen, Germany\\}

\author{Hubert Ebert}
\affiliation{%
  Department  Chemie,  Physikalische  Chemie,  Universit\"at  M\"unchen,
  Butenandtstr. 5-13, 81377 M\"unchen, Germany\\}

\date{\today}

\maketitle

\section{1. Computational details}
All calculations have been performed self-consistently and fully relativistically within the four component Dirac formalism using the coherent potential approximation (CPA) alloy theory as implemented in the Munich SPR-KKR program package \cite{EKM11,SPR-KKR6.3_2}. The treatment of disorder via the CPA was already shown to be a powerful and accurate tool, fully comparable to recent supercell calculations that have to take the average over many configurations \cite{DPM+14,DBEM16}. Doing so, would however give no direct access to the in-plane resistivity anisotropy as it is possible within the presented approach. Furthermore, the CPA allows for an investigation of isovalent doped Ru-122 which is not accessible by e.g. a virtual crystal approximation (VCA). Furthermore, the Dirac formalism implies by construction a full inclusion of effects induced by spin-orbit coupling, which was recently shown to be important for the iron pnictides \cite{BEL+16}. The crystal structure is based on the ortho\-rhombic 4-Fe unit cell in its experimental magnetic configuration \cite{HQB+08}, implying antiferromagnetic chains along the $a$ and $c$ axes and ferromagnetic chains along the $b$ axis. The magnetic moments are oriented in-plane along the $a$ axis. The lattice parameters where chosen according to experimental X-ray data and using a linear interpolation to account for the influence of substitution based on available data \cite{RTJ+08,RPTJ08,SJM+08,TNK+10}. 
More details on the procedure can be found in a previous publication \cite{DPM+14}. For all electronic structure calculations the local density approximation (LDA) exchange-correlation potential was applied, using the parameterization of Vosko, Wilk and Nusair \cite{VWN80}. The longitudinal resistivity was calculated using the Kubo-Greenwood equation within KKR-CPA \cite{But85,BEWV94}, for the symmetric part of the conductivity tensor \eqref{eq_res_Greenwood}: 
\begin{equation}
 \sigma_{\mu\mu} \; = \; \frac{\hbar}{\pi V} \;\text{Tr} \langle \hat j_\mu \text{Im} G^+ \hat j_\mu \text{Im}G^+ \rangle_c \;,\label{eq_res_Greenwood}
\end{equation}
with $G^+$ being the retarded Green's functions, $V$ the cell volume and $\hat j_\mu$ the $\mu$ component of the electric current operator. In the relativistic formulation the latter reads $\hat j_\mu = - |e|c  \alpha_\mu$, with the elementary charge $e$, the speed of light $c$ and $\alpha_\mu$ being one of the $4 \times 4$ Dirac $\alpha$ matrices. The brackets $\langle \; \rangle_c$ indicate a configurational average with respect to an alloy in terms of the CPA. 

\section{2. Vertex corrections}

\begin{figure}[tbh]
{\includegraphics[clip]{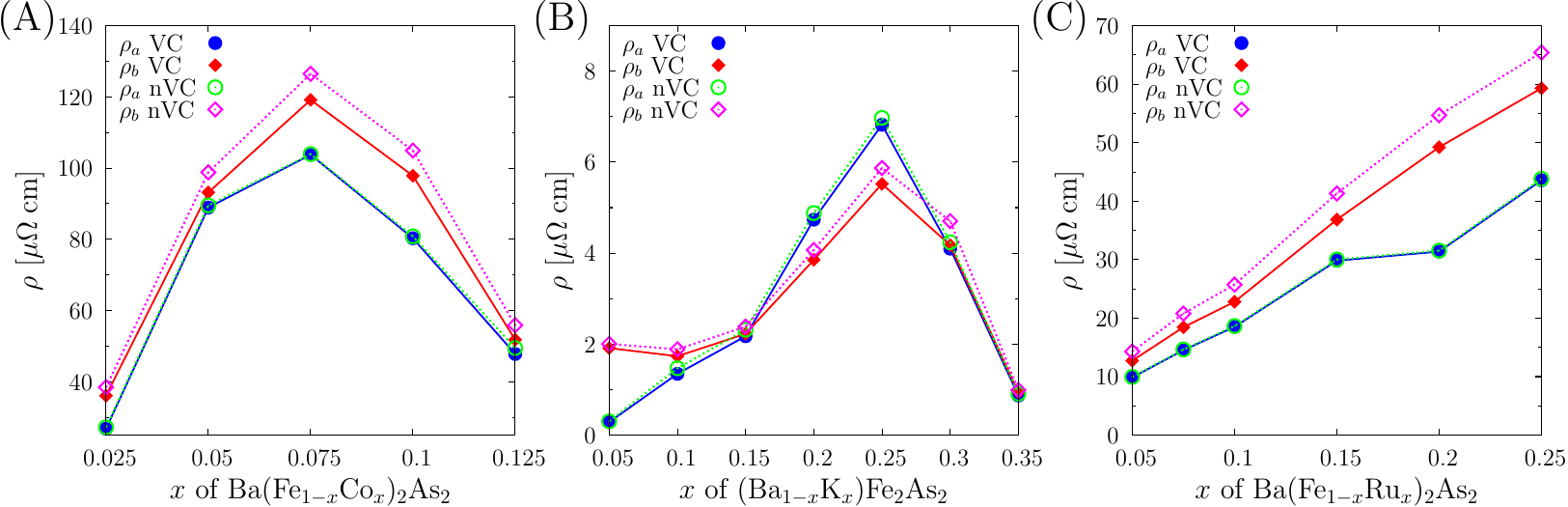}} 
\caption{In-plane resistivity of the AFM state for (A) Co-122, for (B) K-122 and for (C) Ru-122 depending on the substitution level $x$. The results are split into $\rho_a$ and $\rho_b$ either including vertex corrections (VC) with solid lines or without any vertex corrections (nVC) shown in dashed lines.\vspace*{-0.2cm}}\label{Fig_SVC}
\end{figure}

The CPA vertex corrections \cite{But85,BEWV94} are included in all resistivity calculations, ensuring proper averaging of the product of two Green's functions in equation \eqref{eq_res_Greenwood}. Still, the overall impact of the vertex corrections is comparably small (in the order of at most 7~\%) which indicates that incoherent scattering processes are less important. This is in agreement with having mainly $d$ character for the electrons at the Fermi level. Furthermore, this means that it is a reasonable approximation to look at the effective single particle Bloch spectral function in order to investigate this resistivity anisotropy \cite{Vel69}. However, it is interesting to note that the vertex corrections itself are also highly anisotropic as can be seen in Fig.~\ref{Fig_SVC}. Here, $\rho_a$ and $\rho_b$ of the AFM state are shown for all investigated compounds, including vertex corrections (VC) and also without vertex corrections (nVC). It is obvious that the impact of vertex corrections on $\rho_a$ is almost zero in all cases while the influence on $\rho_b$ is especially for Co-122 and Ru-122 much more pronounced. This finding is consistent with a stronger impact of disorder on the AFM coupling along $b$.

\section{3. Lifshitz transition of the parent compound}
\begin{figure*}[tb]
{\includegraphics[clip]{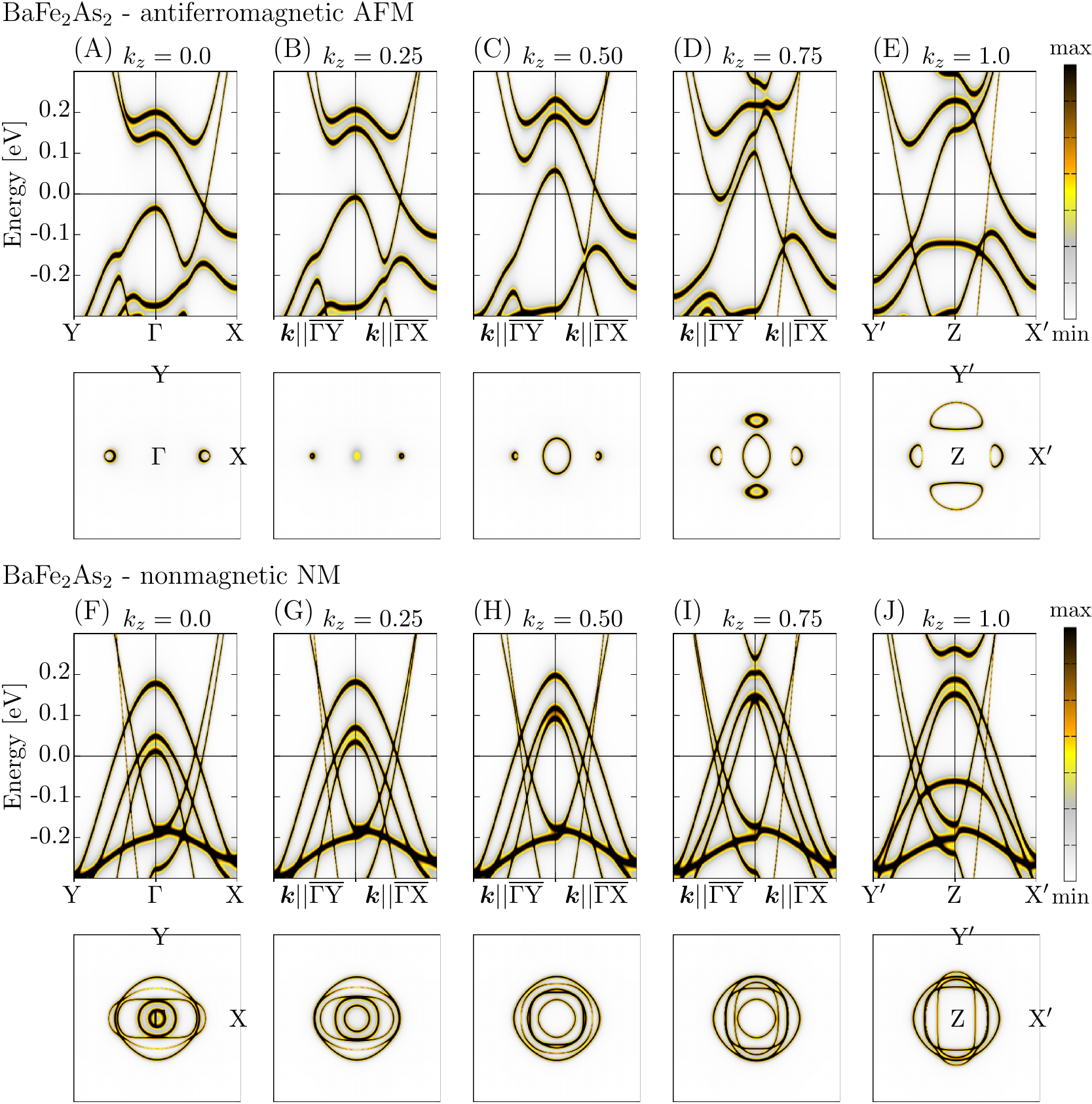}} 
\caption{Bloch spectral functions of orthorhombic \ce{BaFe2As2} for different values of $k_z$ in $\frac{2\pi}{c}$, with $\Gamma$Y and $\Gamma$X corresponding to paths parallel to the $b$ and $a$ axes, respectively. Shown for (A - E) the AFM state and for (F - J) the nonmagnetic case. The upper panel presents the band structure and the lower panel shows the corresponding cut of the Fermi surface. In order to allow direct comparison, all images are shown for the 4-Fe unit cell with a back-folding of bands onto $\Gamma$.\vspace*{-0.0cm}}\label{Fig_SBand}
\end{figure*}

Lifshitz transitions are characterized as a topological change of the Fermi surface.\cite{Lif60}  There are several kinds of Lifshitz transitions discussed for the iron pnictides and these are of significant importance for the underlying physics and superconductivity.\cite{LKF+10,MIM+10,LPD+11,KJ14} The topological change of the Fermi surface for the \ce{BaFe2As2} mother compound when going from the nonmagnetic to the antiferromagnetic state can be also understood in terms of a Lifshitz transition, introducing strong anisotropy into the AFM electronic structure.\cite{DPM+14,DBEM16} The corresponding electronic structure for \ce{BaFe2As2} is shown in Fig.~\ref{Fig_SBand} for the (A - E) AFM state and the (F - J) nonmagnetic case. For comparison, all calculations are presented with respect to the same orthorhombic 4-Fe unit cell which implies a back-folding of bands into a smaller Brillouin zone, compared to the 2-Fe cell. See also Ref. \cite{YLA+09a} for further information on the magnetic Brillouin zone in \ce{BaFe2As2}. The upper rows show the band structure for different values of $k_z$, while the lower rows depict the respective Fermi surface cuts. As in the main paper, the paths $\Gamma$Y and $\Gamma$X correspond to paths in $\bv k$-space which are parallel the $b$ and $a$ axes, respectively.

Going from a 2-Fe unit cell to the 4-Fe cell implies that the hole pockets and the electron pockets of the 2-Fe cell are back folded onto each other in one $\Gamma$ point as can be seen for the nonmagnetic band structure of a 4-Fe unit cell in  Fig.~\ref{Fig_SBand} (F - J). An additionally emerging antiferromagnetic order leads to a hybridization of these hole- and electron pockets as shown in Fig.~\ref{Fig_SBand} (A - E), which was discussed in detail by e.g. Liu \etal\;\cite{LKF+10} in terms of a Lifshitz transition. It is possible to discuss the emergence of a non-zero \rA~already at this point, although earlier attempts were less fruitful.\cite{YLC+11,LMI+15} The reason for this apparent incompatibility is that the iron pnictides are clearly 3D materials whose bands show typically a strong $k_z$ dispersion, which is often not sufficiently accounted for.\cite{DBB+16} This becomes obvious in Fig.~\ref{Fig_SBand} where especially in the AFM case a strong electronic anisotropy emerges which does significantly depend on the value of $k_z$. For the nonmagnetic case in Fig.~\ref{Fig_SBand} (F - J) the corresponding electronic structure is, when considering the $k_z$ dispersion, more or less isotropic. The situation is completely different for the AFM state in Fig.~\ref{Fig_SBand} (A - E). With a similar argumentation as used in the main paper one can qualitatively expect a \rA~with a higher $\rho_b$ along the path $\Gamma$Y for e.g. $k_z = 0.0$ and 0.75. The hybridization does lead first of all to the opening of a band gap only along the $\Gamma$Y path for $k_z = 0.0$. For  $k_z = 0.75$ a minimum of an electron pocket at $E_F$ leads to a minimum of the quasiparticle velocity, increasing the resistivity. Using an even higher resolution in the $k_z$ dispersion (45 different values were calculated) shows clearly that these two distinct features emerge only along $\Gamma$Y and never along $\Gamma$X. This can already qualitatively explain the occurrence of \rA~in the undoped \ce{BaFe2As2}. This effect increases with substitution, as disorder-induced anisotropic band broadening leads to a simultaneous reduction of the quasiparticle lifetime as discussed in the main paper. Electron or hole doping do further move the Fermi level and can either increase or decrease the anisotropy, as was shown in Fig.~4. 

\vspace*{-0.0cm}